\documentstyle[aps,preprint]{revtex}

\def\prl{{\sl Phys. Rev. Lett.}\ }
\def\PRL{{\sl Phys. Rev. Lett.}\ }
\def\ds{\displaystyle}
\def\pr {{\sl Phys. Rev.}\ }

\def\sss                               {
     \scriptscriptstyle                       }
\def\raise                             {
     {\sss +}                            }
\def\spin                            {
     \sigma                             }
\def\k                               {
     k                            }
\def\kdag{K^\raise}

\def\cspindag                            {
     {\bf d}_{\spin}^{\raise}              }
\def\ckspindag                            {
     {\bf c}_{\k\spin}^{\raise}              }

\begin{document}
\draft

\title{Composite Edge States in the $\nu=2/3$ Fractional Quantum Hall Regime}
\bigskip
\author{Yigal Meir}
\address{Physics Department, University of California,Santa Barbara, CA 93106}
\bigskip
\maketitle
\begin{abstract}
A generalized  $\nu=2/3$ state, which unifies the edge-state pictures
of MacDonald and of Beenakker is presented and studied in detail. Using
an exact relation between correlation functions of this state and
those of the Laughlin $\nu=1/3$ wave function, the correlation functions
of the $\nu=2/3$ state are determined via a classical Monte Carlo calculation,
for systems up to $50$ electrons. It is found that as a function of the
slope of the confining potential there is
a sharp transition of the ground state from one description to the other.
The experimental implications are discussed.
\end{abstract}
\pacs{72.20.My, 73.40.Kp, 73.50.Jt\\ \\ \\ \ \ \ \ \ \ \ \ \ \ Submitted to
Physical Review Letters}
\leftline{}
\end{document}
\newpage

The prolific research on the fractional quantum Hall effect in the last decade
has led to a very good understanding of the bulk properties of the fractional
quantum Hall liquid \cite{review}.
The nature of the edge states in this regime, however, is far from being
 well
understood \cite{wen}. Particularly intriguing is the situation for $\nu=2/3$,
where
several theories have been proposed to describe the edge states.
 One picture, due to MacDonald \cite{macdonald}, is
based upon a wavefunction proposed by Girvin \cite{girvin}, which, due to
particle-hole symmetry, consists of droplet of
holes in the $\nu=1/3$ Laughlin-state\cite{laughlin}
 superimposed on a droplet of electrons
in the $\nu=1$ state.
 This wavefunction has indeed been shown to be an excellent description
of the exact ground-state for a system with a small number of electrons
\cite{mac1,jari,yoshi}.
This $\nu=2/3$ wavefunction supports two different edges\cite{macdonald},
one at the edge of the hole droplet (of charge $q=-e/3$),
and the other at edge of the $\nu=1$ electron droplet (of charge $q=e$).
On the other hand, a very different edge structure was suggested
by Beenakker \cite{beenfrac}, and elaborated on by Chklovskii et
al. \cite{chklovskii} in a more general context,  who
 argued  that for a smooth enough potential
an incompressible  $\nu=1/3$
state will nucleate near the edges of the system, leading
again to two edge branches, but this time of charge $q=e/3$ each
 \cite{more}.

Recently it was  argued \cite{jari} that tunneling into a $q=\pm e/3$ edge
state will reduce the tunneling amplitude by a factor of $1/N$ relative to the
integer case.
Hence, tunneling measurements through a small system
in the fractional quantum Hall
 regime offer the exciting possibility of directly probing the composition
of the edge structure of the system. At zero or low magnetic fields
the conductance consists of a series of well separated peaks
\cite{meirav}, each corresponding to an electron tunneling through
a specific state. In the first scenario, where a single edge state
carries a fractional
charge, one would expect that half of the peaks will be suppressed, giving
a clear signature of the composition of the edge states. In the second
scenario, where both edges carry a fractional charge, all the peaks would
be suppressed, resulting in a very low conductance signal.
Until now there has been no quantitative theoretical understanding of
the experimental circumstances require to explore each scenario.

In this work we study quantitatively the nature of the ground state and
the corresponding edge states in the $\nu=2/3$ regime. Generalizing the Girvin
wavefunction to incorporate the possibility of a $\nu=1/3$ strip near the
edge of the sample, the correlation functions in this
generalized state are expressed exactly in terms of correlation functions
calculated in the $\nu=1/3$
Laughlin wavefunction. Using the mapping into
a classical one-component two-dimensional plasma \cite{laughlin} we
calculate those  correlation functions
using classical Monte Carlo \cite{metro} for up to $50$ electrons.
The resulting $\nu=2/3$ correlation functions enable us to calculate the
energy of the state for arbitrary electron-electron interactions and
confining potential. We find that as a function of the slope of
the confining potential,
the ground state makes a sharp transition from
the Girvin form  to the Beenakker form.
This calculation suggests that for heterostructures where the gates are not
too far from the two-dimensional electron gas, the suppression of
half of the peaks, in the first scenario above, should
be observable. In addition, information about the actual distance
between the two edges, which is a crucial ingredient of recent
edge state theories \cite{wen}, is obtained.

The ground state of $N$ electrons in a radially symmetric
system in the $\nu=1/3$ fractional quantum
Hall regime can be approximated very well by the Laughlin wave function
\cite{laughlin},
\begin{equation}
\Psi^{(\sss 1/3)}\left(z_1,...,z_N\right) =
\prod_{i<j}\,(z_i-z_j)^3\ e^{-\sum_i|z_i|^2/4} \ \
 \hat{=}\sum_{\{i_1,\cdots,i_N\}} C^{\sss (N)}_{\{i_1,\cdots,i_N\}}\
 {\bf a}^{\raise}_{i_1}\, \cdots\,{\bf a}^{\raise}_{i_N}\ |\,0> \ \ ,
\label{eq:laughlin}
\end{equation}
where $z_i$ denotes the complex coordinates of the $i$-th particle in the
plane,
and all lengths are expressed in units of the magnetic length,
$\ell_H\equiv \sqrt{\hbar c/eH}$. $\hat{=}$ denotes a second-quantization
representation, where ${\bf a}^{\raise}_{n}$ creates an electron in
 a first Landau-level state of angular momentum $n$, described by the
single-particle wavefunction $\phi_n(z)=z^n\exp(-|z|^2/4)/\sqrt(2\pi 2^n n!)$.
The sum is over all permutations of $N$ distinct integers which sum up
to the total angular momentum
$3N(N-1)/2$, and the $C^{(N)}_{\{i_1,\cdots,i_N\}}$ can, in principle,
be obtained by expanding the first product. In the second-quantization
representation, the particle-hole symmetric wave function, introduced
by Girvin to describe the $\nu=2/3$ state \cite{girvin}, is
\begin{equation}
\Psi^{(\sss 2/3;N_h)}\left(z_1,...,z_N\right)  \
\hat{=}\sum_{\{i_1,\cdots,i_{\sss N_h}\}} C^{\sss (N_h)}_{\{i_1,\cdots,i_{\sss
N_h}\}}\
 {\bf a}_{i_1}\, \cdots\,{\bf a}_{i_{\sss N_h}}\
{\bf a}^{\raise}_1\, \cdots\,{\bf a}^{\raise}_{\sss N+N_h}\ |\,0> \ \ .
\label{eq:girvin}
\end{equation}
The yet undetermined number of holes, $N_h$,
must be chosen to minimize the energy.

In order to allow for the possibility of a $\nu=1/3$ state nucleating
along the edge of the sample, we start with
the Laughlin wavefunction with an inside hole of size $L$\cite{laughlin},
\begin{equation}
\Psi^{(\sss L;1/3)}\left(z_1,...,z_N\right) =
\prod_{i} z_i^{\sss L} \prod_{i<j}\,(z_i-z_j)^3\ e^{-\sum_i|z_i|^2/4} \ \
 \hat{=}\sum_{\{i_1,\cdots,i_{\sss N}\}} D^{\sss (N;L)}_{\{i_1,\cdots,i_N\}}\
 {\bf a}^{\raise}_{i_1+{\sss L}}\,
\cdots\,{\bf a}^{\raise}_{i_{\sss N}+{\sss L}}\ |\,0> \ \ ,
\label{eq:shifted}
\end{equation}
where the sum is over the same sets as in (\ref{eq:laughlin}),
and $D^{\sss (N;L)}$
can again, in principle, be evaluated by expanding the products.
With the above  wavefunction describing electrons with $\nu=1/3$
correlations along the edge of the sample, we write a generalized
$\nu=2/3$ state,
\begin{eqnarray}
\Psi^{(\sss 2/3;N_h,L,N_2)}\left(z_1,...,z_N\right)&\nonumber \\
\hat{=} \ds{\sum_{\{i_1,\cdots,i_{_{\sss N_h}}\}}
\sum_{\{j_1,\cdots,j_{_{\sss N_2}}\}}}
& C^{\sss (N_h)}_{\{i_1,\cdots,i_{_{\sss N_h}}\}}\
D^{\sss (N_2;L)}_{\{j_1,\cdots,j_{_{\sss N_2}}\}}\
 {\bf a}_{i_1}\, \cdots\,{\bf a}_{i_{\sss N_h}}\
{\bf a}^{\raise}_1\, \cdots\,{\bf a}^{\raise}_{\sss N_1+N_h}\
{\bf a}^{\raise}_{j_1+{\sss L}}\,
\cdots\,{\bf a}^{\raise}_{j_{_{\sss N_2}}\!\!+{\sss L}}\ |\,0> \ \ .
\label{eq:tt}
\end{eqnarray}
This wave function is schematically depicted in Fig.1. It depends on
three integer parameters. Out of the $N$ electrons,  $N_1$ are described by
the Girvin wavefunction (\ref{eq:girvin}), with
$N_h$ holes.
The remaining $N_2=N-N_1$ electrons nucleate into
a $\nu=1/3$ strip along the edge of the sample, with minimal
angular momentum $L$ ($L>N_1+N_h$).

Our task now is to find the set of parameters that minimizes the energy
for a given  confining potential and interactions.
 To
this end we need to calculate the one-particle and two-particle
correlation functions for this state.
In principle, of course, if one
can obtain the coefficients $C^{\sss (N_h)}_{\{i_1,\cdots,i_{_{\sss N_h}}\}}$
and $D^{\sss (N_2;L)}_{\{j_1,\cdots,j_{_{\sss N_2}}\}}$, all correlation
functions
for the $\nu=2/3$ state (Eq.(\ref{eq:tt}))
should be easily evaluated. This, however, can only be achieved
for a system of a very small number ($N\leq6$) of particles \cite{jari,expand}.
Correlations functions for the $\nu=1/3$
Laughlin-like wave function (\ref{eq:shifted}), on the other hand,
 can be easily calculated for a large number of particles, using
a mapping into a classical statistical problem \cite{laughlin}.
Such a mapping, however, is not possible for the Girvin wavefunction
(\ref{eq:girvin}), or the generalized wavefunction (\ref{eq:tt}).

The most important step in this work is expressing the
correlation functions for the generalized $\nu=2/3$ wavefunction
in terms of correlation functions for the Laughlin-like wavefunctions
for $\nu=1/3$. Using the explicit form of this wavefunction (\ref{eq:tt}),
we find,
\newpage
\begin{eqnarray}
\rho_1^{(\sss 2/3;N;N_h,L,N_2)}(r) &=& \rho_1^{(\sss 1;N_1)}(r) -
\rho_1^{(\sss 1/3;N_h)}(r) + \rho_1^{(\sss 1/3;N_2;L)}(r)
\nonumber \\
\rho_2^{(\sss 2/3;N;N_h,L,N_2)}(r_1,r_2) &= &\rho_2^{(\sss 1;N_1)}(r_1,r_2)
+ \rho_2^{(\sss 1/3;N_h)}(r_1,r_2) + \rho_2^{(\sss 1/3;N_2;L)}(r_1,r_2)
\nonumber \\
-&&\!\!\!\!\!\!\!\!\!\!\!\!\rho_1^{(\sss 1;N_1)}(r_1)\rho_1^{(\sss
1/3;N_h)}(r_2) -
 \rho_1^{(\sss 1;N_1)}(r_2)\rho_1^{(\sss 1/3;N_h)}(r_1) +
\rho_1^{(\sss 1;N_1)}(r_1)\rho_1^{(\sss 1/3;N_2;L)}(r_2) \nonumber \\
+&&
\!\!\!\!\!\!\!\!\!\!\!\!\rho_1^{(\sss 1;N_1)}(r_2)\rho_1^{(\sss
1/3;N_2;L)}(r_1)
-\rho_1^{(\sss 1/3;N_h)}(r_1)\rho_1^{(\sss 1/3;N_2;L)}(r_2) -
\rho_1^{(\sss 1/3;N_h)}(r_2)\rho_1^{(\sss 1/3;N_2;L)}(r_1)
\nonumber \\
+& \ \ 2 Re& \left[\ds{\sum_{i=0}^{3(N_h-1)}}\ \
\ds{\sum_{j=0}^{N_1-1}}<\!n_i\!>_{\!\!_{1/3}}^{\!\!^{(N_h)}} \phi_i^*(r_1)
\phi_i(r_2)
\phi_j^*(r_2) \phi_j(r_1)\right]
\nonumber \\
-& \ \ 2 Re& \left[\ds{\sum_{i=L}^{L+3(N_2-1)}}\ \
\ds{\sum_{j=0}^{N_1-1}}<\!n_i\!>_{\!\!_{1/3}}^{\!\!^{(N_2;L)}} \phi_i^*(r_1)
\phi_i(r_2)
\phi_j^*(r_2) \phi_j(r_1)\right]
\ \ \ \ \ \ \ \ \ \ \ \ \ \ \  \ \ \ \ \ \ \ \ \ \
\nonumber \\
+& \ \ 2 Re &\left[\ds{\sum_{i=L}^{L+3(N_2-1)}}\ \
\ds{\sum_{j=0}^{3(N_h-1)}}<\!n_i\!>_{\!\!_{1/3}}^{\!\!^{(N_2;L)}}
<\!n_j\!>_{\!\!_{1/3}}^{\!\!^{(N_h)}}
 \phi_i^*(r_1) \phi_i(r_2)
\phi_j^*(r_2) \phi_j(r_1)\right]  .
\label{eq:mapping}
\end{eqnarray}
The single-particle distribution function $\rho_1$, normalized
such that its integral is $N$, is simply
expressed as the sum of the three distribution functions of the $N_1$ electrons
in the $\nu=1$ state ($\rho_1^{(1;N-N_2)}$),  that of the $N_2$ electrons
in the strip of $\nu=1/3$ state ($\rho_1^{(1/3;N_2;L)}$), and (minus) that of
 the $N_h$ holes in the $\nu=1/3$ state,
($\rho_1^{(1/3;N_h)}$). The two-particle correlation function, $\rho_2$, here
normalized to $N(N-1)$,
 is far more complicated. Nevertheless, the various terms contributing
to the resulting interaction energy have straightforward interpretations.
The first three terms describe the contribution to the interaction energy
from interactions within the three different components. The next six
terms describe the direct (Hartree) interactions between the three components.
The last three nontrivial terms correspond to the exchange and correlation
interactions between the different components.

Eq. (\ref{eq:mapping}) enables us to express the one- and two-particle
correlation functions in terms of quantities evaluated for the $\nu=1/3$
states (Eq.\ref{eq:laughlin} and Eq.\ref{eq:shifted}). We calculate
these correlation functions for the $\nu=1/3$ state
using classical Monte Carlo
method \cite{metro,datta}.
 In addition we have to calculate
$<\!n_i\!>_{1/3}$, the average occupation of the $i$-th state. This
quantity is somewhat more complicated to calculate, and so far it has
 been calculated with only a partial success \cite{mac3}. In this work we
calculate
it using the relation,
\newpage
\begin{eqnarray}
|\Psi^{(\sss 1/3;N)}\left(z_1,...,z_N\right)|^2 &=& \prod_{i=2}^N |z_1-z_i|^6
e^{-|z_1|^2/2} |\Psi^{(\sss 1/3;N-1)}\left(z_2,...,z_N\right)|^2  \nonumber \\
&=&
\sum_{i=0}^{3(N-1)} |\phi_i(z_1)|^2 F_i(z_2,\cdots,z_n)
|\Psi^{(\sss 1/3;N-1)}\left(z_2,\cdots,z_N\right)|^2 .
\label{nm}
\end{eqnarray}
Thus $<\!n_i\!>_{1/3}$ which are
the  coefficients of $|\phi_i|^2$ in the expansion of the density can be
directly evaluated by numerically calculating the averages of
  the functions $F_i$
in the $N-1$-particle system. The one-particle distribution function
deduced from the occupations has been compared with the direct
Monte Carlo calculation of the  distribution function and an
excellent agreement was found.

Having calculated the correlation function for the generalized
$\nu=2/3$ state (\ref{eq:tt}), its energy can easily be evaluated for
any choice of interactions and confining potential. In order to be as
close to the experimental situation as possible, we have chosen a
Coulomb interaction, $U(r)=e/\epsilon r$. The confining potential
was chosen
to rise linearly from
zero to its maximum value, $S$, over a length $d$.
The position of the
 midheight of the potential step is fixed so the filling factor
is $2/3$.
As discussed below,
the physically relevant parameter will be the slope of the potential,
$S/d$.
$d$ is determined
experimentally by the distance of the gates from the two-dimensional
electron gas, while $S$ is determined by the amount of voltage applied
 to the gate, as seen by the electrons in the 2d gas. For typical
Ga-As samples, the gates are 120-200nm from the 2d gas, which corresponds
to 8-12 magnetic lengths.
The interaction energy $e/\epsilon \ell_H$ is
typically 5 meV, while the boundary potential seen by the electrons
is tens of meV \cite{frank}. Here we will express all energies in units
of $e/\epsilon \ell_H$. The claculation are done for
up to $50$ electrons, which is a typical
number in an experimental quantum dot\cite{meirav}.

In Fig.2a we plot the number of holes, $N_h$, which minimizes the energy
for a step potential ($d=0$), for two values of $S=3$ and $S=5$. For
a step potential, the ground state usually involves $N_2=0$ electrons in the
$\nu=1/3$ strip, so it is of the Girvin type (\ref{eq:girvin}).
The number of holes in the ground state is determined by the competition
between the two contributions to the energy: the larger the number of
holes, the more uniform the density, and the lower the
interaction energy. On the other hand, the larger the number of holes,
the larger the potential energy.
Thus,
as the potential becomes smoother,
the number of holes may increase and a strip of electrons in the $\nu=1/3$
state may form near the edge.

As can be seen from the figure, the number of holes in the ground state
scales as $N/2$, which shows the region of density different from $\nu=2/3$
is independent of $N$, namely an edge effect. From Fig.2a one can obtain
the physical distance between the edges for large enough system, as a function
of the confining potential. We find that this distance changes from
$\sim 1.5\ell_H$ to $\sim 2.5\ell_H$ when $S$ changes from $3$ to $10$.
Thus, unlike the case for slowly varying confining potential\cite{chklovskii}
 one cannot consider those edges as isolated from each other, and
any theory should include interactions and mixing of those states.

Since the number of holes is an integer, the number of
holes will change, on average, every other time an electron is added to
the system. This is the source of the prediction \cite{jari} that half
of the peaks for  tunneling into a $\nu=2/3$ droplet
will be suppressed. As the present calculation cannot
produce the tunneling amplitudes exactly, we estimate them by
their upper limit,
the average occupation of the angular momentum state
the electron tunnels to.
In the inset of Fig.2a we plot this occupation as a function of $N$.
The suppression of more than
half of the peaks is clearly observed, with the right
power-law dependence on the electron number. Interestingly, the
calculation suggests that sometimes the ground states of $N$ and $N+1$
electrons differ by two holes. It remains to be seen if this is a real
effect, which will result in a more dramatic reduction of the peak
amplitude.

In any real system the potential will rise over a finite length scale, $d$.
We have studied the nature of the ground state as a function of $d$,
and we found that for a given electron number $N$, and potential height $S$,
there will be a transition from the ground state being of
the Girvin type (\ref{eq:girvin}) to a state which includes electrons
nucleating at the edge of the sample in the $\nu=1/3$ state. By varying
$S$ it is found the transition occurs at the same ratio of $S/d$, namely
at a given slope of the confining potential.
For example,
 for $30$
electrons the ground state evolves smoothly from
$N_h=9$ and $N_2=0$, for $d=0$, to  $N_h=5$ and $N_2=0$ for $S/d\simeq1$,
and then it changes abruptly
 to $N_h=15$, $N_2=2$ and $L=N_1+N_h+1$
 (Fig. 2b). Thus the two edges,
of the electron droplet and the hole droplet  suddenly merge, and
a $\nu=1/3$ strip forms, signaling a transition from the
Girvin-MacDonald picture to the
picture presented
by Beenakker \cite{beenfrac} and Chklovskii et al. \cite{chklovskii}.
This
strip moves away from the edge of the electron droplet ($L=N_1+N_h+1$)
as $S/d$ decreases. For example, for $S/d=0.6$ the ground state corresponds to
$N_h=15$, $N_2=2$ and $L-(N_1+N_h)=20$ (Fig. 2b).
For $40$ electrons  one can actually see
two transitions. For $S/d\simeq1$ the ground state changes from $N_h=12$
and $N_2=0$ to   $N_h=19$ and $N_2=0$, namely it is still described by
Eq.(\ref{eq:girvin}), but the two edges have merged, while for
$S/d\simeq1.4$ nucleation first occurs with $N_h=18$ and $N_2=5$.
This intermediate regime where the two edges merged may suggest
a possible description in terms of a single $\nu=2/3$ edge \cite{fisher}.
As the slope of the potential in
experimental systems\cite{meirav} is if the order of
$1.2-3$ $e^2/\epsilon \ell_H^2$ \cite{frank},
we predict that the suppression of half of the tunneling peaks should
be observable in quantum dots in present high mobility structures.
 The closer the gates to the 2d gas, the better the chances
 of seeing that effect.  In addition,  it is predicted
that as a function of the voltage applied to the gates, (which
changes the slope of the effective potential), the tunneling
peak structure will change abruptly as this transition occurs. For high
voltages half of the peaks appearing in the $\nu=1$ regime will be
suppressed in the $\nu=2/3$ regime, while
for lower voltages, as extensive tunneling into the $\nu=1/3$ state
will occur, most or all of the peaks will be suppressed.

In conclusion, using an exact expression for the generalized $\nu=2/3$-state
 correlation
functions in terms of the $\nu=1/3$ ones, we have been able to study
quantitatively systems of relatively large number of electrons ($N\leq 50$).
Consequently we predict a transition in the nature of the
ground state of the system as a function of the slope of the
confining potential and discuss its experimental
manifestation.
It is hoped that this work will stimulate more experiments
in this direction.

I thank M. P. A. Fisher, W. Kohn and X.-G. Wen for useful
discussions, and N. S. Wingreen for a critical reading of this manuscript.
This  work was supported by NSF grant no. NSF-DMR90-01502  and by the
NSF Science and Technology Center for Quantized Electronic Structures,
Grant no. DMR 91-20007. The numerical calculations in this work have
been performed on the San Diego Supercomputer CRAY-YMP. Additional
calculations have been performed using resources of the Cornell Theory
Center, which receives major funding from the National Science Foundation
and IBM Corporation, with additional support from New York State Science
and Technology Foundation and members of the Corporate Research
Institute.
\vskip 8truecm
\leftline{\sl Figure Captions}

1. Schematic representation of the generalized $\nu=2/3$ state (\ref{eq:tt}).
 Out
 of the $N$ electrons,
$N_1$ are described by the Girvin wavefunction, consisting
 of $N_h$ holes in the $\nu=1/3$ state
in the background of $N_1+N_h$ particles in
 the $\nu=1$ state (\ref{eq:girvin}).
The remaining $N_2$ electrons nucleate into a $\nu=1/3$ strip
 along the edge of the sample.

2. (a) The number of holes that minimizes the energy of the Girvin wavefunction
 (\ref{eq:girvin}) for two different sizes of step potentials.
The straight lines correspond to $N_h\propto N/2$,
leading to an N-independent edge size.
 The inset shows the tunneling amplitude, as estimated from
 the occupation of the relevant state, as a function of $N$, on a log-log plot.
 The suppression of at least half of the peaks is evident and agrees very well
 with the theoretically predicted $1/N$ dependence \cite{jari}.
 (b) The density profile of the ground states of $N=30$ electrons for $3$
different
  potential slopes. The existence of a $\nu=1$ region is evident for the
highest
 slope, while for the other two slopes an incompressible $\nu=1/3$ strip is
formed
 along the edge.


\begin{references}
\bibitem[11]{more} In fact, as the potential becomes smoother and
smoother, more incompressible states are expected to nucleate along
the edge of the sample. Such very smooth potentials are not dealt with
in this work.
\end{references}
\end{document}